\documentclass[epj]{svjour}
\usepackage{graphics}
\begin{document}
\title{Experimental Evidence for the Pentaquark}
\author{Daniel S. Carman
\thanks{email: carman@ohio.edu}%
}                     
%
%
\institute{Ohio University, Athens, OH, 45701}
\date{Received: date / Revised version: date}
%
\abstract{
The present experimental evidence for the existence of light pentaquarks is 
reviewed, including both positive and null results.  I also discuss the CLAS 
experiments at Jefferson Laboratory that are forthcoming in the near future 
to address questions regarding existence, mass, width, and other quantum 
numbers of these five-quark baryon states.
} 
\maketitle
\section{Introduction}
\label{intro}

The body of new experimental results indicating the possible existence of 
pentaquark baryons represents one of the most exciting topics to arise in 
nuclear physics in the past decade.  To date there have been nearly a 
dozen experiments employing various beams and targets that have announced 
evidence for a narrow state (called $\Theta^+$) with a mass of about 
1540~MeV/c$^2$ and a width of $<$20~MeV.  There are also a similar number 
of high-energy experiments that see no evidence for such a state.  
Additionally one experiment has announced evidence for a narrow cascade 
state (called $\Xi_5$) at a mass of 1862~MeV/c$^2$ with a width of $<$20~MeV.

Speculation regarding baryon states whose minimal quark content cannot be
obtained with only three valence quarks has appeared in the literature
for more than 30 years, initially in connection with $K-N$ scattering
data~\cite{golowich}.  These states, initially referred to as $Z^*$ resonances, 
were included in the Particle Data Group (PDG) listings until 1986, when they 
were dropped due to a lack of convincing experimental evidence.  However 
a recent theoretical paper by Diakonov~\cite{diakonov}, which predicted an 
exotic baryon with minimal quark content $uudd\bar{s}$ at a mass of 
1540~MeV/c$^2$ with a width of 15~MeV in a chiral soliton framework, 
reawakened the field and has led to a new round of experimental searches.  The
isoquartet of $\Xi_5$ states along with the $\Theta^+$ are believed to be part
of an anti-decuplet of pentaquark baryon states with spin 1/2 and positive
parity. Of important note is that most of these new searches come from mining 
existing data that were taken for other purposes.  This is important to appreciate, 
especially given the relatively low statistical significance of the published 
positive results (in the range from 4-7$\sigma$).

The experimental concerns regarding the $\Theta^+$ include not only the 
statistical significance of the observed peaks, but also the apparent 
discrepancy between the masses measured in the $nK^+$ and $pK^0$ isospin 
decay modes.  The differences among the determined $\Theta^+$ mass values must
be addressed to ensure there are no experimental problems.  However, the mass
difference could be attributed to different initial and/or final state 
interactions or different interference effects in the two decay modes.  For 
the $\Xi_5$ baryon, only one experiment has been able to see any hint of a 
signal.  Of concern in each of these analyses is the meaning and importance 
of different analysis cuts used to enhance the signal.  However, these cuts
could also provide important clues regarding the production dynamics of these 
exotic states.  Providing a definite answer to the question of existence or 
non-existence of the $\Theta^+$, $\Xi_5$, and other five-quark baryons is of 
critical importance in this field in order to better understand QCD and the 
structure of hadronic matter.

In this talk I focus on the present experiment evidence, including both positive
and null results, for the low-mass exotic five-quark $\Theta^+$ and $\Xi_5$ 
baryon states.  I then focus on several of the key theoretical issues related to 
these states.  Finally I discuss the second generation, high statistics experiments 
that are being planned with CLAS at Jefferson Laboratory that will be crucial in 
providing definitive answers regarding the existence and the nature of the 
pentaquark states.

\begin{table*}[htbp]
\begin{center}
\begin{tabular} {|c|l|c|c|c|c|} \hline \hline
{\bf Expt.} & {\bf ~~~Reaction} & {\bf Mass} (MeV/c$^2$) & {\bf Width} $\Gamma$ (MeV) & {\bf Significance} & {\bf Ref.}\\ \hline
{\tt LEPS}  & $\gamma C \to K^+K^-X$           & $1540 \pm 10$& $< 25$     & $4.6\sigma$    & \cite{nakano} \\ \hline
{\tt DIANA} & $K^+ Xe \to K_s^0 p X$           & $1539 \pm 2$ & $< 9$      & $4.4\sigma$    & \cite{barmin} \\ \hline
{\tt CLAS-d}& $\gamma d \to K^+ K^- p (n)$     & $1542 \pm 5$ & $< 21$     & $5.2\sigma$    & \cite{stepanyan} \\ \hline
{\tt SAPHIR}& $\gamma p \to K^+ K_s^0 (n)$     & $1540 \pm 6$ & $< 25$     & $4.8\sigma$    & \cite{saphir} \\ \hline
{\tt CLAS-p}& $\gamma p \to K^+ \pi^+ \pi^- (n)$ & $1555 \pm 10$& $< 26$     & $7.8\sigma$    & \cite{kubarovsky} \\ \hline
{\tt ITEP}  & $\nu A \to K_s^0 p X$            & $1533 \pm 5$ & $< 20$     & $6.7\sigma$    & \cite{neutrino} \\ \hline
{\tt SVD}   & $p+A \to p K_s^0 X$              & $1526 \pm 3$ & $< 24$     & $5.6\sigma$    & \cite{svd}  \\ \hline
{\tt HERMES}& $e^+ d \to K_s^0 p X$            & $1526 \pm 3$ & $13 \pm 9$ & $\sim 5\sigma$ & \cite{hermes} \\ \hline
{\tt ZEUS}  & $e p \to K_s^0 p X$              & $1522 \pm 3$ & $8 \pm 4$  & $\sim 5\sigma$ & \cite{zeus} \\ \hline
{\tt COSY}  & $pp \to K_s^0 p \Sigma^+$        & $1530 \pm 5$ & $< 18$     & $\sim 5\sigma$ & \cite{cosy-tof} \\ \hline \hline
\end{tabular}
\end{center}
\caption{Overview of the positive $\Theta^+$ pentaquark experiments.  The column labeled
``Significance'' lists the quoted statistical significance as explained in the text.}
\label{pos_ex}
\end{table*}

\section{Positive Experimental Searches}
\label{posex}

At the current time at least 10 different experiments have announced a 
positive result for the $\Theta^+$ baryon.  A compilation of the mass and 
decay width from these experiments is presented in Table~\ref{pos_ex}.  
The estimated statistical significance quoted is given by $N_s/\sqrt{N_b}$, 
where $N_s$ is the number of counts within $\pm$2$\sigma$ of the mean for the 
$\Theta^+$ peak from a Gaussian fit and $N_b$ is the corresponding number of 
background events.  Each of the experiments has relatively low statistics and 
the backgrounds beneath the peaks are not completely understood, which 
obviously affects the quoted statistical significance.  The first group of 
experiments from the LEPS, DIANA, CLAS, and SAPHIR collaborations have searched 
for evidence of the $\Theta^+$ using different beams, energies, detector 
configurations, and assumptions regarding the reaction mechanism.  As each was 
carried out using data collected for other purposes, each is statistically and 
systematically limited.  In this section I focus on selected aspects of these 
different measurements.

The LEPS experiment at SPring-8 employed a photon beam with tagged energies
between 1.5 and 2.4~GeV.  The $\gamma n \to K^+ K^- n$ reaction was studied
on carbon from a scintillator target~\cite{nakano}.  Cuts were made to
remove the dominant $\phi \to K^+ K^-$ decay channel and backgrounds
from $\gamma p \to K^+ K^- p$.  Corrections were made for the Fermi momentum
of the target nucleon, leading to a $\Theta^+$ signal at a mass of 1540~MeV/c$^2$
with 19 counts above a background of 17 counts.  This analysis assumed that
the remainder of the nucleus acted as a spectator to the interaction.  This work 
has since been followed up with a dedicated run on a deuterium target.  Preliminary 
results announced at the $N^*2004$ conference indicate a consistent result with
roughly a factor of four increase in statistics~\cite{nakano2}.

The DIANA experiment at ITEP employed a $K^+$ beam of 750~MeV/c incident 
on a xenon-filled bubble chamber~\cite{barmin}.  The analysis focussed on 
$K^+ Xe \to K_s^0 p X$ charge-exchange events with $K_s^0 \to \pi^+ \pi^-$,
where the strangeness of the final state $K_s^0$ is known as the $K^+$ has
strangeness +1.  The analysis employed cuts to remove low momentum $p$ and $K_s^0$ 
tracks and cuts on the $K_s^0 p$ opening angle to reduce effects of rescattering
within the nucleus (which is claimed to be an important effect).
The final spectrum shows 29 events in the $\Theta^+$ peak (concentrated
into a single bin) over a background of 44 events.

\begin{figure}
\vspace{8.5cm}
\includegraphics{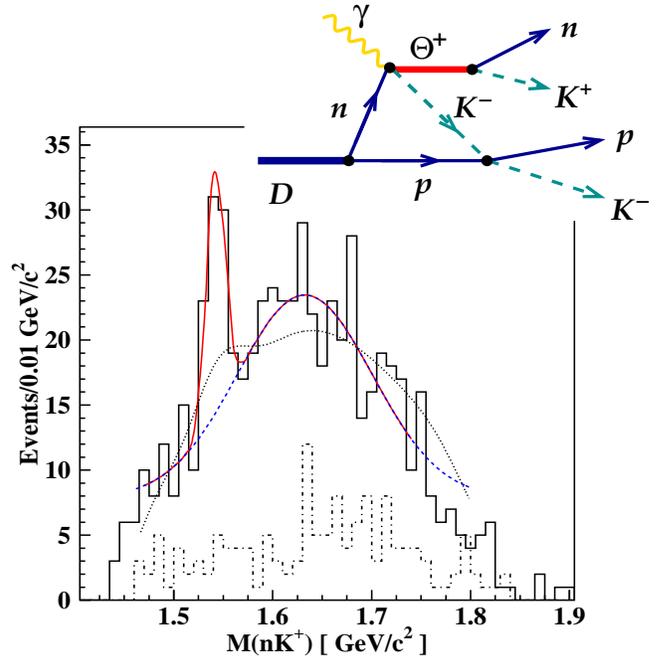}
\caption{Published CLAS-d result~\cite{stepanyan} for the reaction 
$\gamma d \to K^+K^-p(n)$ showing evidence for the $\Theta^+$ peak.  The
curves include two different assumptions for the underlying background. The 
inset represents one possible rescattering diagram.}
\label{clas-d}
\end{figure}

\begin{figure}
\vspace{8.0cm}
\includegraphics{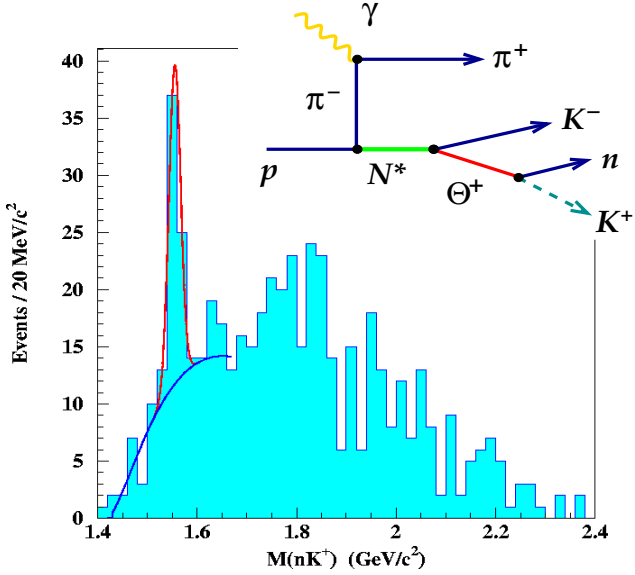}
\caption{Published CLAS-p result~\cite{kubarovsky} for the reaction 
$\gamma p \to K^+ \pi^+ \pi^- (n)$ showing evidence for the $\Theta^+$ peak.
The inset represents the assumed reaction mechanism.}
\label{clas-p}
\end{figure}

The CLAS experiment on deuterium (CLAS-d), $\gamma d \to K^+ K^- p (n)$ 
(the neutron was not detected), was the first exclusive reaction for the 
$\Theta^+$~\cite{stepanyan}.  One of the strengths of an exclusive analysis
is the reduction of background channels and the fact that no Fermi momentum 
correction is required. This experiment used a tagged-photon beam with energies 
from 1.5 to 3.1~GeV. The analysis required a rather complicated secondary 
rescattering mechanism of the final state $K^-$ and $p$ to increase the acceptance
for these particles in CLAS (see Fig.~\ref{clas-d}).  As a result of this assumption 
(which was found to be required to explain $\Lambda(1520)$ production in the same 
final state), the shape of the background is difficult to estimate and may
include some degree of kinematic reflections~\cite{dzierba}.  This experiment
claims 42 counts in its $\Theta^+$ spectrum at a mass of 1542~MeV/c$^2$ and
a width of 21~MeV.  Another CLAS photoproduction experiment was on a proton
target (CLAS-p)~\cite{kubarovsky} $\gamma p \to K^+ \pi^+ \pi^- (n)$ (the neutron
was not detected), and claimed evidence for a $\Theta^+$ with an impressive 
statistical significance of 7.8~$\sigma$ (see Fig.~\ref{clas-p}).  This experiment 
employed tagged photons from 3.0 to 5.5~GeV.  The final event sample is enriched 
by requiring angular cuts to select events with a forward-going $\pi^+$ 
($\cos \theta_\pi^{CM} > 0.8$) and a backward-going $K^+$ ($\cos \theta_K^{CM} < 0.6$).  
These cuts were designed to remove the dominant background from meson production 
reactions and enhance the $s$-channel production mode for the $\Theta^+$.  These
data were examined by a partial wave analysis where the amplitudes of each
partial wave were fit over the full angular coverage of CLAS, hence fixing
the background under the $\Theta^+$ peak.  The analysis claims 41 events in
the $\Theta^+$ peak at a mass of 1555~MeV/c$^2$, which is $\sim$15~MeV higher
than the CLAS-d result.

The SAPHIR experiment was the first to present evidence for the $\Theta^+$
from a proton target from the exclusive reaction $\gamma p \to K^+ K_s^0 (n)$
\cite{saphir} (the neutron was not detected).  They employed a tagged-photon 
beam with energy between 0.8 and 2.6~GeV.  To enhance the $\Theta^+$ signal, 
they employed a cut on forward-going kaons ($\cos \theta_K^{CM} > 0.5$)
and observed 63 events in their $\Theta^+$ peak. The large cross section
that they estimated from their data conflicted with the data from CLAS for
the same reaction.  A re-analysis of the SAPHIR data now suggests a smaller
cross section, but the result is still under investigation~\cite{saphir2}.

Following these papers, several other experiments measured the $pK_s^0$ 
invariant mass via inclusive reactions on nuclei and claimed to see evidence 
for the $\Theta^+$.  One of the experiments collected data from 120k 
40-100~GeV $\nu$ and $\bar{\nu}$ charged-current events on hydrogen, deuterium, 
and neon-filled bubble chambers~\cite{neutrino}.  Two others experiments
from the HERA accelerator employed electroproduction data.  The result from
HERMES~\cite{hermes} employed a fixed-target $e^+d$ experiment at 28~GeV.  
The result from ZEUS~\cite{zeus} used $e-p$ collisions at 
$\sqrt{s} \approx$310~GeV. The SVD experiment focussed on $p+A$ collisions
at 70~GeV/c at the IHEP accelerator requiring $\cos \theta_{pK_s^0} \ge 0$
\cite{svd}. Of importance in these experiments is that the $K_s^0$ is a mixture 
of both strangeness +1 and -1 eigenstates, so the invariant mass spectra could 
include both $\Sigma^{*+}$ and $\Theta^+$ peaks.  However, it has been reasoned
that a peak at a $pK_s^0$ mass where there are no known $\Sigma^{*+}$ states 
can be interpreted as evidence for the strangeness +1 $\Theta^+$.  It is also 
interesting that these four experiments each report a $\Theta^+$ mass which is 
about 10~MeV below those experiments reporting a $\Theta^+$ from $nK^+$ 
reconstructions.  Furthermore, most of the null experiments for the $\Theta^+$ come 
from analyses using the same method of measuring the $pK_s^0$ invariant mass.  The 
inherent weakness in not knowing the strangeness of a particle, coupled with the 
uncertainty in the associated backgrounds, clearly raises an important concern with 
these experiments. To reduce these doubts, it is important to confirm the $pK_s^0$ 
peak in experiments where the final state strangeness is cleanly tagged.

\begin{figure}
\vspace{5.7cm}
\includegraphics{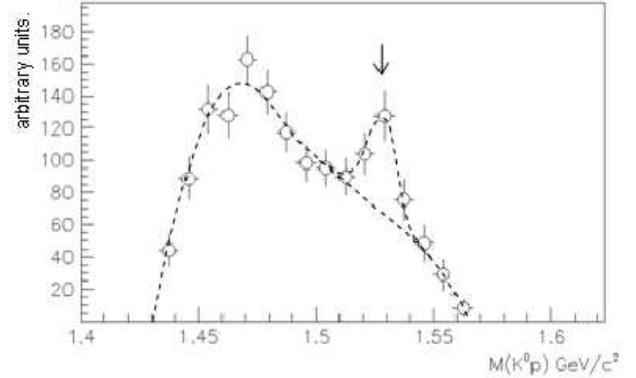}
\caption{Published COSY-TOF result~\cite{cosy-tof} for the reaction 
$pp \to \Sigma^+ K_s^0p$ showing evidence for the $\Theta^+$ peak.}
\label{cosy}
\end{figure}

Another more recent result with strong evidence for the $\Theta^+$ is from
the COSY-TOF experiment~\cite{cosy-tof} which employed the exclusive hadronic 
reaction $pp \to \Sigma^+ K_s^0 p$ reaction at $p_p$ = 2.95~GeV (see 
Fig.~\ref{cosy}).  Here the strangeness of the $K_s^0 p$ combination is tagged 
by the accompanying final state $\Sigma^+$.  Particle identification is done 
entirely by geometric reconstruction, which is quite accurate for this near 
threshold reaction.  This analysis yielded a consistent result from the
2000 and 2002 data sets for the $\Theta^+$ mass of 1530~MeV/c$^2$ and a cross
section of 0.4~$\mu$b.  This collaboration has plans to acquire additional
data with improved detector resolution within the next year.

\begin{figure}
\vspace{7.5cm}
\includegraphics{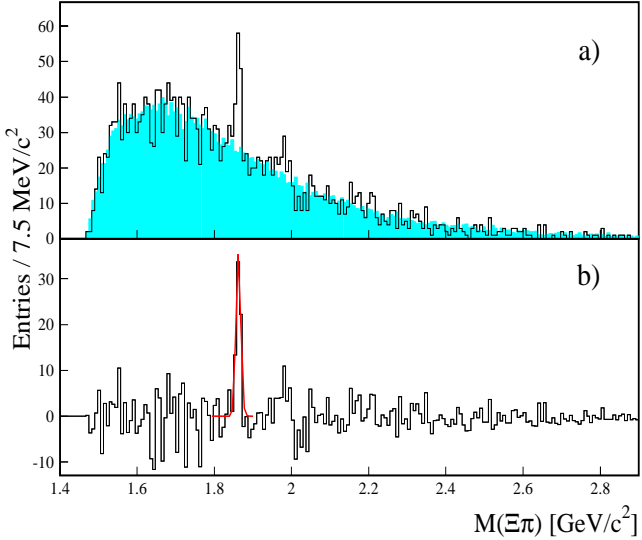}
\caption{Published NA49 result~\cite{na49} for (a). the sum of the $\Xi^{-}\pi^{-}$, 
$\Xi^{-}\pi^{+}$, $\overline{\Xi}^{+}\pi^{-}$ and $\overline{\Xi}^{+}\pi^{+}$ 
invariant mass spectra. The shaded histogram shows the normalized mixed-event 
background. (b). Background subtracted spectrum with the Gaussian fit to the peak.}
\label{na49}
\end{figure}

There is one other relevant and potentially exciting analysis.  The NA49 
experiment at CERN claims to see evidence for the exotic $\Xi_5^{- -}$
pentaquark state decaying into $\Xi^-\pi^-$ with a mass of 1862~MeV/c$^2$ 
and a width of 18~MeV via $pp$ reactions at $\sqrt{s}$=17.2~GeV~\cite{na49}.
This analysis finds 38 $\Xi_5^{- -}$ candidates above a background of 43
events (see Fig.~\ref{na49}).  This mass is in disagreement with the chiral
soliton predictions of Diakonov~\cite{diakonov} and the diquark model 
predictions of Jaffe and Wilczek~\cite{jaffe}.  No other experiments that 
have searched for this exotic state have been able to confirm this result.  
There is also additional controversy from within the NA49 collaboration itself 
that leads to doubt about this result~\cite{fischer}.  Thus confirmation of 
this signal awaits and is in fact crucial to understand the nature and 
dynamical underpinnings of the $\Theta^+$ itself.

\section{Null Experimental Searches}
\label{nullex}

\begin{figure}
\vspace{8.0cm}
\includegraphics{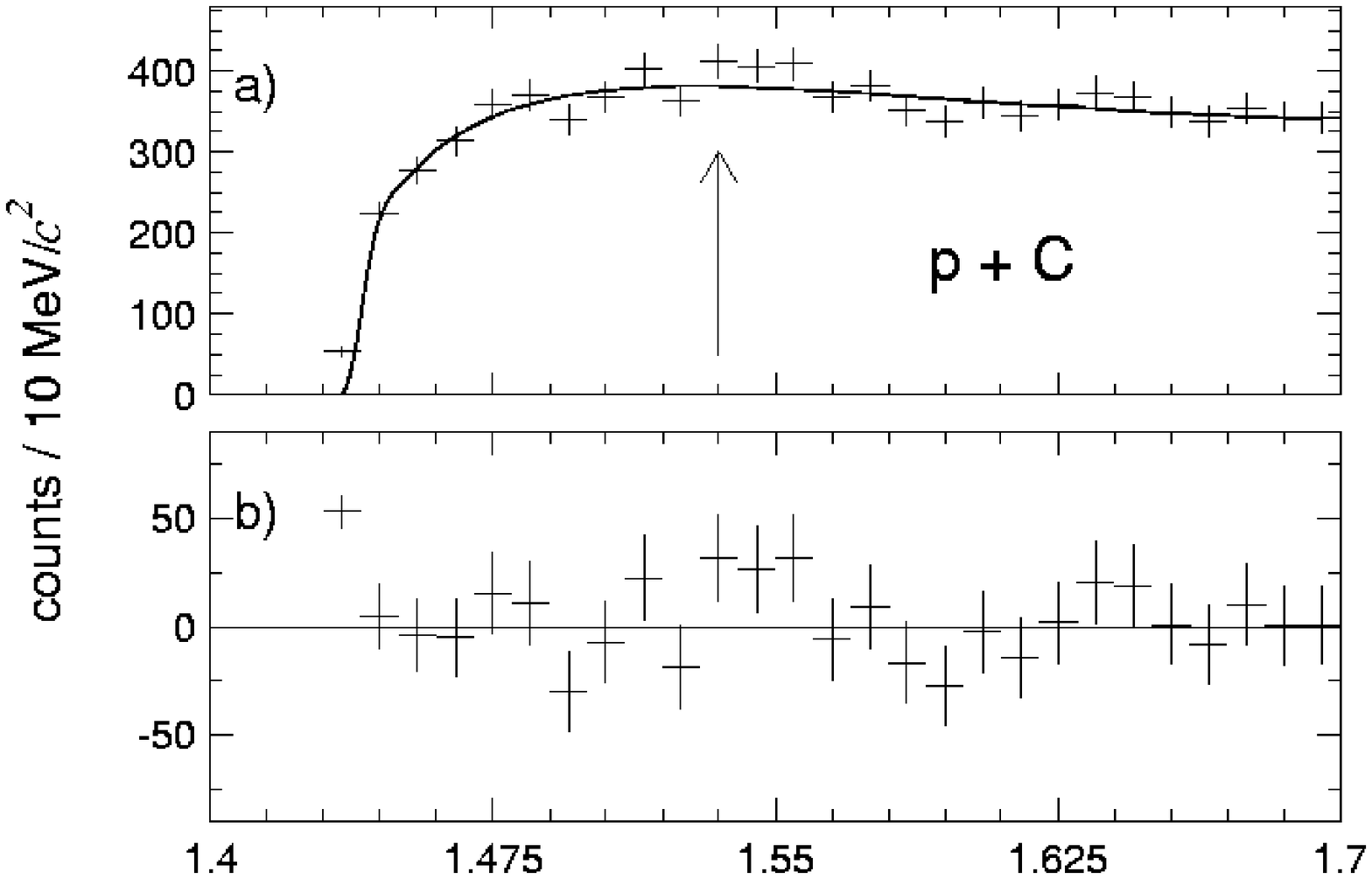}
\includegraphics{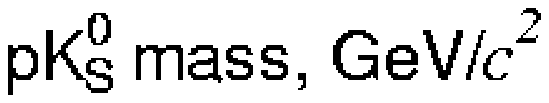}
\caption{Published HERA-B result~\cite{herab} for the $pK_s^0$ invariant
mass from $p+C$ collisions at 920~GeV.  (a). Shows background determined 
from event mixing as the solid line, (b). Background subtracted data.}
\label{herab}
\end{figure}

During the course of the past year a number of null results for the $\Theta^+$ 
and $\Xi_5$ states have been presented at various conferences.  These experiments 
have been compiled in Table~\ref{null_ex}, and mainly report null results in high 
energy reactions (see Fig.~\ref{herab} for the HERA-B result).  
These experiments see many of the well known hadronic resonances, but are mainly 
from high multiplicity, inclusive analyses (with the exception of the Fermilab E690 
experiment).  Because of the difficulty in detecting neutrons in these experiments, 
they have typically focussed on reconstructions of the $pK_s^0$ invariant mass, 
just as in the HERMES and ZEUS experiments.

\begin{table*}[htbp]
\begin{center}
\begin{tabular}{|c|c|c|} \hline
{\bf Expt.} & {\bf Reaction}                                             & {\bf Ref.} \\ \hline \hline
HERA-B      & $p+A$ at 920 GeV, $\sqrt{s}$=41.6~GeV                      & \cite{herab} \\ \hline
\rule{0mm}{3mm}
BES         & $e^+e^- \to \psi(2s) {\rm~or~} J/\psi \to \Theta^+ \bar{\Theta}^-$; B.R.$<$10$^{-5}$ & \cite{bes} \\ \hline
CDF         & $p\bar{p}$ at low and high $p_t$                           & \cite{cdf} \\ \hline
BaBar       & $e^+ e^-$ at $\sqrt{s}=11$~GeV                             & \cite{qnp} \\ \hline
ALEPH,DELPHI,OPAL & $e^+e^- \to Z \to q\bar{q}$ at $\sqrt{s}=91$~GeV     & \cite{qnp}\\ \hline
FNAL E690   & $pp \to pK^-\pi^+ \Theta^+$ at 800~GeV; exclusive reaction & \cite{e690} \\ \hline
HyperCP     & Mixed beam $\pi,K,p$ at 120-250~GeV                        & \cite{hypercp}\\ \hline
\rule{0mm}{3mm}
PHENIX      & $d+Au$ at $\sqrt{s}=200$~GeV $\bar{\Theta}^- \to \bar{n}K^-$ & \cite{qnp} \\ \hline \hline
\end{tabular}
\end{center}
\caption{Overview of null pentaquark experiments. Of the entries in this list, only
the E690 experiment from Fermilab is an exclusive reaction.}
\label{null_ex}
\end{table*}

Initially one might expect that if the $\Theta^+$ exists, it should be produced
in both high-energy electroproduction and in high-energy hadron collisions.
However, reconstruction of the $\Theta^+$ (and the $\Xi_5$ as well) must also
be intimately connected with understanding the production mechanism for
such states, as well as in understanding the underlying background processes.
One of the most obvious differences between experiments reporting evidence
for the $\Theta^+$ and experiments that report a null result is the different
kinematics in the experiments.  The differences between the different
experiments must surely provide strong clues as to why exclusive measurements
at medium energy show a potential $\Theta^+$ peak, whereas this signal is
not equally apparent in high-energy inclusive measurements.  One potentially
relevant argument may be that the reaction mechanism is indeed through excitation
of an intermediate crypto-exotic $N^*$ state as hinted at by the CLAS-p
analysis~\cite{kubarovsky}.

Whatever the final explanation is, it is clear that more data and theoretical
input is required to address all of the questions raised.  From my viewpoint
I am entirely convinced that the peaks seen in the positive pentaquark experiments
in Table~\ref{pos_ex} are not statistical fluctuations.  However, it is not yet
fully apparent what has given rise to these signatures, whether they are due
to kinematic reflections, detector inefficiencies, final state interactions,
interference effects, or true pentaquark states.

\section{$\Theta^+$ Width}
\label{theory}

One of the most intriguing aspects of the $\Theta^+$ is its apparent
narrow width.  Within any model of five quarks moving independently
in an effective mean field potential, the pentaquark state would
fall apart rapidly with a width of $\sim$100~MeV.  Thus if the $\Theta^+$
truly exists, it must be exotic dynamically, as well as in its quantum
numbers.

The chiral soliton model of Diakonov~\cite{diakonov} predicts a narrow 
width for the $\Theta^+$ based on the underlying symmetries of the model.  
However the predicted width has significant model dependence.  Quark models 
can also obtain widths on the order of $\le$10~MeV if there are non-trivial 
quark rearrangements within the $\Theta^+$ wavefunction.  For example the 
diquark model of Jaffe and Wilczek~\cite{jaffe} proposes that the $\Theta^+$ 
is composed of a pair of scalar $ud$ diquarks and a strange anti-quark.  
Another example is the diquark-triquark model of Karliner and Lipkin
\cite{lipkin} where two color non-singlet clusters are kept apart by the 
$p$-wave angular momentum barrier.

The recent data for the $\Theta^+$ suggest that the state is so narrow
that most experiments can only set an upper bound on its width of
$\sim$20~MeV, consistent with experimental resolutions.  Recent
detailed analysis of the $K-N$ scattering database indicates that
the width of the $\Theta^+$ pentaquark would have to be less than
several MeV to have escaped detection~\cite{trilling,arndt}.  On the other
hand, the $K-N$ database is quite sparse in this interesting energy
range.  In fact an important future need is high resolution measurements 
of the $K^+d \to K_s^0 p (p)$ charge-exchange cross section for $p_{K^+}$ 
in the range of 400-500~MeV/c as a function of the final state $K_s^0p$ mass.

To date two experiments have measured a width that is more than an upper limit.  
The HERMES experiment has quoted a width for the $\Theta^+$ of 13$\pm$9~MeV
\cite{hermes}, which is larger than their experimental resolution of 4-6~MeV.  
The ZEUS experiment measured an intrinsic width for the $\Theta^+$ of 8$\pm$4~MeV
\cite{zeus}, which is above their experimental resolution of 2~MeV.  These widths, 
however, depend on the assumptions for the background shape.

If the $\Theta^+$ (and the $\Xi_5$ as well) really is narrow as a few MeV, 
then the present chiral soliton and quark models will have a difficult time
providing a complete dynamical explanation.  Another important question regards 
the $\Theta^+$ production mechanism if its width is indeed $\sim$1~MeV.  For 
photoproduction, the most straightforward calculation of the production mechanism 
is via $t$-channel kaon exchange, where the width is related to the coupling
constant $g_{K\Theta N}$ and the phase space for production.  If the width is 1~MeV, 
then this diagram is suppressed, although $K^*$ exchange is still possible
\cite{hicks}.  Another intriguing possibility is production through a crypto-exotic 
$N^*$ resonance, first suggested by the CLAS proton data~\cite{kubarovsky}.  If 
this production mode is dominant, then this might provide a natural explanation 
for some of the null results.  However, there are still many more questions than 
answers and we must await further studies to make progress.

\section{Second Generation Experiments}
\label{secex}

As noted earlier, nearly all of the pentaquark analysis results obtained to 
date have resulted from data sets initially acquired for other experiments.  
In most cases this has resulted in data that are not entirely sufficient to 
exclude that the observed signals are due to statistical fluctuations, 
kinematic reflections, or other subtle experimental artifacts.  The limited 
statistics do not permit detailed checks of systematic dependencies.

\begin{table}[htbp]
\begin{center}
\begin{tabular}{|c|c|c|c|} \hline
Run & Energy & Reactions & Running \\ \hline
g10 & 3.8 GeV & $\gamma d \to \Theta^+ K^- p$   & Completed \\
    &         & $\gamma d \to \Theta^+ \Lambda$ & \\ 
    &         & $\gamma n \to \Theta^+ K^-$     & \\ \hline
\rule{0mm}{3mm}
g11 & 4.0 GeV & $\gamma p \to \Theta^+ \bar{K}^0$ & Completed \\
    &         & $\gamma p \to \Theta^+ K^- \pi^+$ & \\ \hline
eg3 & 5.7 GeV & $\gamma_v p \to \Xi_5^{- -} X$ & Dec. 2004 \\
    &         & $\gamma_v p \to \Xi_5^+ X$ & \\ \hline
g12 & 5.7 GeV & $\gamma p \to \Theta^+ K^- \pi^+$ & 2005 \\
    &         & $\gamma p \to \Theta^+ \bar{K}^0$ &      \\
    &         & $\gamma p \to K^+ K^+ \Xi_5^-$ & \\ \hline        
\end{tabular}
\end{center}
\caption{Overview of new pentaquark experiments at CLAS.}
\label{overview}
\end{table}

Given the importance of coming to a definitive conclusion regarding
the $\Theta^+$ and other pentaquark baryons, the CLAS experiment
at Jefferson Laboratory has dedicated the majority of its beam time
in 2004-2005 to focus on providing answers.  An overview of the
approved experiments is contained in Table~\ref{overview}.

The g10 experiment which took data in the spring of 2004 will
focus on $\Theta^+$ production via $\gamma d \to p K^- \Theta^+$,
$\gamma d \to K^- \Theta^+$, and $\gamma d \to \Lambda \Theta^+$.  
The g11 experiment, which completed data taking immediately after the g10 
experiment, will focus on the reactions $\gamma p \to \Theta^+ \bar{K}^0$ 
and $\gamma p \to \Theta^+ K^- \pi^+$.  Both g10 and g11 will look
for $\Theta^+$ decays into $nK^+$ and $pK^0$.  These experiments
have collected more than an order of magnitude more statistics than
the existing published CLAS data~\cite{stepanyan,kubarovsky}.  Both 
experiments have implemented procedures to reduce the uncertainty
in the reconstructed $M(NK)$ mass spectra to the few MeV level, which
will be important to study any possible mass differences between the
$\Theta^+$ decay modes.

The anti-decuplet of five-quark states within the chiral soliton or quark 
models contains an isoquartet of $\Xi_5$ states, two of which are manifestly
exotic, the $\Xi_5^{- -}$ ($ddss\bar{u}$) and $\Xi_5^+$ ($uuss\bar{d}$).
Evidence for such states has been seen in only one experiment to date
\cite{na49}.  It is imperative to either confirm or refute these data with 
a high statistics analysis.  At CLAS, two new experiments are planning to 
take data toward this end during the next year.

The eg3 experiment will use an untagged photon beam to study reactions
producing $\Xi_5^{- -}$ and $\Xi_5^+$ pentaquark states by reconstructing
the final state from the cascade decay chain.  A crucial technique to
reduce the combinatoric backgrounds in this experiment is to reconstruct
the different detached vertices with CLAS.  The second CLAS $\Xi_5$ 
experiment is part of the g12 experiment, which will focus on the production 
reaction instead of the decay process as in eg3.  Here the $\Xi_5^-$ will be 
studied via the missing mass in the reaction $\gamma p \to K^+ K^+ X$.  CLAS 
has already successfully analyzed the $\Xi$(1320) ground state with this 
technique in existing lower energy data.

\section{Summary}
\label{summ}

With roughly equal amounts of positive and null evidence for the exotic 
$\Theta^+$ and $\Xi_5$ pentaquarks from a variety of experiments under a
broad variety of conditions, one cannot conclude whether these states and 
other non-$qqq$ baryon states exist.  Furthermore, the theoretical 
difficulties in explaining a width perhaps as small as 1~MeV, suggest that if
these states do exist, they are very usual.  At this point in time the guidance 
from lattice gauge theory is not particular elucidating as the available
calculations come to widely different conclusions.  It is clear that the burden 
to unequivocally prove or disprove the existence of these baryon states is an 
experimental task.  Presently the PDG has listed the $\Theta^+$ as a 3-star 
resonance, although many feel that this assignment is premature given all of 
the apparently conflicting data.

At the present time several second generation high statistics detected
pentaquark experiments have either completed data taking or are scheduled at 
various facilities including CLAS.  If either or both of the $\Theta^+$ and 
$\Xi_5$ pentaquark states can be established experimentally, the new data will 
allow for significant progress on the development of a detailed program of 
pentaquark spectroscopy.  These data can prove invaluable in establishing the 
spin, parity, and isospin quantum numbers of these states, as well as providing 
valuable insight into the associated production mechanisms.  It is also 
essential to study evidence for other spin-orbit partners of these states, as 
well as evidence for the other non-exotic pentaquarks and to understand how
five-quark states couple to and mix with ordinary three-quark states.

\section{Acknowledgements}

I am grateful to my colleagues in the CLAS Collaboration who have contributed
heavily to this area of research.  This talk would not have been possible
without the hard work of many experimentalists and theorists and without the
many interesting interactions I have had with many individuals.  This work
was supported in part by the Department of Energy and the National Science
Foundation.

%
%
%

\end{document}